\documentclass[epj]{svjour}
\usepackage{graphics}
\usepackage{amssymb}
\usepackage{amsmath}
\usepackage[dvips]{epsfig}
\newcommand{\fmn}[2]{\mbox{${\textstyle \frac{#1}{#2}}$}}
\newcommand{\rd}{\mbox{{\rm d}}}
\newcommand{\bmath}[1]{\mbox{\boldmath $#1$}}

\newcommand{\ddpipi}{\mbox{$dd\to\,^{4}\textrm{He}\,\pi\pi$}}
\newcommand{\pdpipi}{\mbox{$pd\to\,^{3}\textrm{He}\,\pi\pi$}}
\newcommand{\ddeta}{\mbox{$dd\to\,^{4}\textrm{He}\,\eta$}}
\newcommand{\pdeta}{\mbox{$pd\to\,^{3}\textrm{He}\,\eta$}}

\begin{document}
\title{Two--pion production in deuteron--deuteron collisions at low energies}
\author{G\"{o}ran F\"{a}ldt\inst{1}$\,$\thanks{\email{goran.faldt@tsl.uu.se}}
\and
Ulla Tengblad\inst{1}$\,$\thanks{\email{ulla.tengblad@tsl.uu.se}}
\and
Colin Wilkin\inst{2}$\,$\thanks{\email{cw@hep.ucl.ac.uk}}}
\institute{
Department of Radiation Sciences, Box 535, S-751 21 Uppsala, Sweden
\and
Department of Physics \& Astronomy, UCL, London WC1E 6BT, UK}
\date{Received: \today / Revised version:}
\abstract{The cross section for the \ddpipi\ reaction is estimated
near threshold in a two--step model where a pion created in a
first interaction produces a second pion in a subsequent
interaction. This approach, which describes well the rates of
$2\pi$ and $\eta$ production in the \pdpipi\ and \ddeta\
reactions, leads to predictions that are much too low compared to
experiment. Alternatives to this and the double--$\Delta$ model
will have to be sought to explain these data.
 }
\maketitle
%
%
\section{Introduction}
Over the last few years there has been increased experimental
interest in double-pion production near threshold in several
hadronic reactions. These include studies in
pion--proton~\cite{CB} and proton--proton collisions~\cite{Heinz},
as well as in the \pdpipi~\cite{MOMO,Andersson,Heinz2} and
\ddpipi~\cite{Pia1,Pia2,Pia3} reactions. For excess energies $Q$
below about 100$\:$MeV one sees no sign of the low mass $s$-wave
$\pi\pi$ enhancement, known as the ABC effect~\cite{ABC}, and the
maxima in the invariant mass distributions tend more to be pushed
to the highest possible values.

Due in part to an isospin filter effect, the most spectacular
manifestation of the ABC is to be found in the case of \ddpipi\
for $Q\approx 200\!-\!300\:$MeV~\cite{Ban73}. These cross section
data, as well those representing the deuteron analysing
powers~\cite{SPESIII}, can be well understood within a model where
there are two independent pion productions, through the $pn\to
d\pi^0$ reaction, with a final state interaction between the two
deuterons to yield the observed $\alpha$--particle~\cite{Anders}.
Since the $pn\to d\pi^0$ amplitudes are dominated by $p$--wave
production, driven by the $\Delta$ isobar, this leads to much
structure in the predictions. Although such double--$\Delta$
effects are generally observed in the medium energy
data~\cite{Ban73,Anders}, there is little evidence of them nearer
to threshold~\cite{Pia1,Pia2,Chapman}. Furthermore, the cross
sections measured at low energies are over an order of magnitude
higher than the predictions of models behaving like the square of
$p$-wave production, where the amplitudes must be proportional to
$Q$.

In an alternative approach to the \pdpipi\ reaction, the low
energy cross sections have been discussed in terms of a two--step
model, where a pion is produced through a $pp\to d\pi^+$ reaction
on the proton in the deuteron, with a further pion being created
in a secondary $\pi^+n\to p\pi^0\pi^0$ reaction~\cite{FGW}. There
are, of course, other contributions related to this through
isospin invariance. Semi-phenomenological models of the $\pi^+n\to
n\pi^0\pi^0$ amplitudes show strong $s$--wave production, behaving
rather like a contact term, plus another contribution involving
the decay chain $N^*(1440)\to \Delta(1232)\,\pi \to
N\pi\pi$~\cite{Oset}. The $s$--wave term is sufficient, in the
two--step model, to lead to reasonable agreement with the
available data on the \pdpipi\ total cross section. Moreover,
combined with $p$--waves required by the decay chain, it
reproduces the shift of the mass spectrum away from the ABC region
towards that of higher missing masses~\cite{MOMO,Andersson}. It is
therefore reasonable to ask whether a similar approach could not
be usefully tried for the low energy \ddpipi\ reaction.

The two--step model with an intermediate
$pd\to\,^{3}\textrm{He}\,\pi^0$ step has in fact been applied
successfully to the production of $\eta$--mesons in the \ddeta\
reaction near threshold~\cite{FW2}, where it reproduces reasonably
well the magnitude of the total cross
section~\cite{Frascaria,Willis}. The approach is here extended in
section~\ref{sec2} to describe the \ddpipi\ reaction, using the
same $\pi N\to\pi\pi N$ amplitudes as those that worked for
\pdpipi. The other element that is crucial for the evaluation of
this model is the cluster decomposition of the $\alpha$--particle
in terms of $^3$He$\,n/\,^3$H$\,p$ constituents. This is discussed
in section~\ref{sec3}, where we rely on the work of the Argonne
group~\cite{VMC}. The results presented in section~\ref{Results}
show that the model is capable of describing the shape of the
$\pi\pi$ effective mass distribution, without the oscillatory
structure predicted by the double--$\Delta$ model~\cite{Anders}.
However, the total cross section estimates fall over an order of
magnitude below the experimental results found at low
energies~\cite{Pia1,Pia2,Chapman}. These data have low statistics
and limited acceptance, though they will be supplemented by more
precise results expected soon from CELSIUS~\cite{Pia3}. Since
neither this nor the double--$\Delta$ model gets even close to the
observed production rates, alternative approaches are necessary.
%
%
\section{The reaction model}
\label{sec2}

\begin{figure}[htb]
\begin{center}
\centerline{\epsfxsize=8cm{\epsfbox{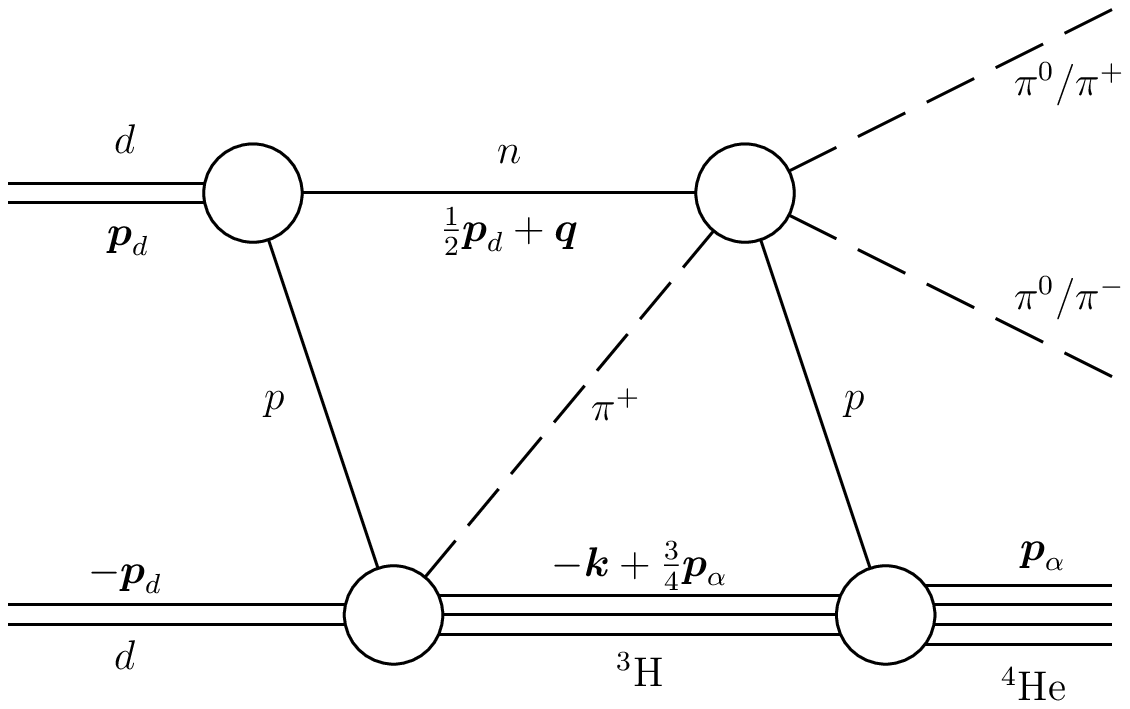}}}
\end{center}
\vspace{-0.2cm} \caption{\label{diagram} Two--step model for the
production of $\pi^+\pi^-$ and $\pi^0\pi^0$ pairs through the
\ddpipi\ reaction. There are also contributions related to this by
isospin in addition to the terms arising from the interchange of
the two deuterons.}
\end{figure}

The two--step model for the \ddpipi\ amplitudes, in terms of those
for $pd\to\,^3\textrm{H}\,\pi^+$ and $\pi^+n\to (\pi\pi)^0p$, is
depicted in Fig.~\ref{diagram}. Contributions involving
intermediate $^3$He and $\pi^0/\pi^-$ are all related to the
results for this diagram through isospin invariance. Due to the
identical nature of the incident deuterons, there is a similar set
of diagrams where the initial production takes place on the upper
deuteron.\vspace{-2mm}

The cross section corresponding to such a diagram has been
evaluated for the $dd\to\alpha\,\eta$ reaction~\cite{FW2} and we
follow closely the techniques used there. The unpolarised \ddpipi\
differential cross section is expressed in terms of the Lorentz
invariant matrix element $\mathcal{M}$ through
\begin{eqnarray}
\rd\sigma&=&\frac{p_{\alpha}}{ 144p_dW^2} \frac{1}{(2\pi)^4}
\sum_{\rm spins} \mid \mathcal{M}\mid^2\, k_{\pi\pi}\,
\rd{}m_{\pi\pi}\,
\rd\Omega_{\alpha}\,\frac{\rd\Omega_{\pi\pi}}{4\pi}\:\cdot
  \label{Start-cross}\nonumber\\
\end{eqnarray}
Here $p_d$ and $p_{\alpha}$ are the initial and final momenta in
the overall cm system where the total energy is $W$. The angles
$\Omega_{\alpha}$ are also in the total cm system, whereas the
$\pi\pi$ relative momentum $k_{\pi\pi}$ and its angles
$\Omega_{\pi\pi}$ are evaluated in the dipion rest frame, where
the total energy is $m_{\pi\pi}$.

The matrix element of Fig.~\ref{diagram} involves the integration
of the pion propagator between the two production vertices over
the two Fermi momenta $\bmath{k}$ and $\bmath{q})$. If initially
we neglect the deuteron $D$--state and the Lorentz boost of the
wave functions, this can be written as
\begin{eqnarray}
\nonumber \mathcal{M}&=&\sqrt{\frac{2}{3 m_p^2}}\
\frac{1}{(2\pi)^3}  \int{\rd^3k }\,{\rd^3q}\,
\frac{m_n}{E_n(\bmath{p}_n)}\frac{m_t}{E_t(\bmath{p}_t)}\\
&&\times\frac{i}{q_{\pi}^2-m_{\pi}^2+i\epsilon}\tilde{\mathcal{M}}_N\:,
\label{full}
\end{eqnarray}
where the particle masses are denoted by $m_i$. The reduced
nuclear matrix element is
\begin{eqnarray}
\tilde{\mathcal{M}}_N&=&\textrm{Tr}\left[
\frac{-1}{\sqrt{2}}\,\bmath{\sigma}\cdot\bmath{\epsilon}_d
\left\{-\mathcal{A}\,\hat{\bmath{p}}_d\cdot
\bmath{\epsilon}_{d\,'} -i\mathcal{B}\,\hat{\bmath{p}}_d\cdot
(\bmath{\epsilon}_{d\,'}
\times\bmath{\sigma})\right\}\right.\nonumber\\
&&\left.\times\ \frac{-1}{\sqrt{2}}\,a(m_{\pi\pi},Q)\,
\bmath{\sigma}\cdot\bmath{p}_{\pi}\right]
\tilde{\varphi}_d(\bmath{q}) \,
\tilde{\psi}^{\dagger}_{\alpha}\,(\bmath{k})\:, \label{reduced}
\end{eqnarray}
where the $(\bmath{\epsilon}_{d},\bmath{\epsilon}_{d\,'})$ are the
polarisation vectors of the two incident deuterons and the
kinematics are defined as in the figure. The $S$-state
momentum--space wave functions for the deuteron and the
triton--proton configuration of the $\alpha$--particle are denoted
by $\tilde{\varphi}_d(\bmath{q})$ and
$\tilde{\psi}_{\alpha}\,(\bmath{k})$ respectively.

In the forward and backward (cm) directions, only two terms are
needed to describe the spin structure of the
$dp\to\,^{3}\textrm{He}\,\pi^0$ amplitude. Using two--component
spinors to denote the $^3$He ($u_h$) and proton ($u_p$), this
reads
\begin{equation}
\mathcal{M}(dp\to\,^{3}\textrm{He}\,\pi^0) =
u^{\,\dagger}_h\left[\mathcal{A}\,\hat{\bmath{p}}_d\cdot
\bmath{\epsilon}_d +i\mathcal{B}\,\hat{\bmath{p}}_d\cdot
(\bmath{\epsilon}_d\times\bmath{\sigma}) \right]u_p\:,
\end{equation}
where $\bmath{p}_d$ and $\bmath{p}_{\pi}$ are the momenta of the
incident deuteron and produced pion respectively. In our
normalisation, the unpolarised differential cross section and
deuteron tensor analysing power $t_{20}$ are given in terms of the
two dimensionless spin amplitudes $\mathcal{A}$ and $\mathcal{B}$
by
\begin{eqnarray}
\nonumber \frac{\rd\sigma}{\rd\Omega} &=&\frac{1}{3(8 \pi
W)^2}\frac{p_{\pi}}{p_d}\Big[\mid\mathcal{A}\mid^2
 +2\mid\mathcal{B}\mid^2 \Big]\\
t_{20}&=&\sqrt{2}\left[\frac{\mid\mathcal{B}\mid^2
 -\mid\mathcal{A}\mid^2}{\mid\mathcal{A}\mid^2
 +2\mid\mathcal{B}\mid^2}\right]\,,
\end{eqnarray}
and these observables have been well measured in collinear
kinematics at Saturne~\cite{Kerboul}.

For deuteron kinetic energies of interest here, the backward
($\theta_{p\pi}=180^{\circ}$) values of $t_{20}$ are strongly
negative, so that $|\mathcal{A}|\gg|\mathcal{B}|$. In the 0.5 --
0.8$\:$GeV range the results may be represented by
\begin{eqnarray}
\nonumber%
|\mathcal{A}|^2&\approx&
-565.6+2318.7T_d-2869.9T_d^2+1122.9T_d^3\hspace{10mm}\\
|\mathcal{B}|^2&\approx&
-197.9+1144.9T_d-2113.0T_d^2+1261.8T_d^3\:,
\end{eqnarray}
where the deuteron kinetic energy $T_d$ is measured in GeV.

The spin structure of the $\pi^-p\to \pi^0\pi^0n$ amplitude is
unique near threshold:
\begin{equation}
\label{e0} M(\pi^-p\to \pi^0\pi^0n)= a(m_{\pi\pi},Q)\,
u_n^{\dagger} \bmath{\sigma}\cdot\bmath{p}_{p}\,u_{p}\:.
\end{equation}
In terms of the amplitude $a$, the unpolarised differential cross
section is
\begin{equation}
\label{e1} \textrm{d}\sigma(\pi^-p\to \pi^0\pi^0n) =
\frac{1}{64\pi^3}\,\frac{p_p\,p_n}{W_{\pi
N}^2}\,|a(m_{\pi\pi},Q)|^2\, k_{\pi\pi}\,\textrm{d}m_{\pi\pi}\:.
\end{equation}
Here $p_p$ and $p_n$ are respectively the initial and final
nucleon momenta, $W_{\pi N}$ the cm energy in the $\pi N$ system,
and $Q=W_{\pi N}-2m_{\pi}-m_N$ the excess energy above the
two--pion threshold.

The low energy data in different isospin channels are well
described by the Valencia model~\cite{Oset} and this allows one to
project out the $I=0$ combination required as input in
equation~(\ref{reduced}). The results can be parameterised as:
\begin{eqnarray}
\nonumber
\lefteqn{\hspace{-5mm}\frac{1}{64\pi^3}\,|a(m_{\pi\pi},Q)|^2 =
(1.092-0.0211Q+0.00015Q^2)}&&\\ \nonumber
&&+(4.18+0.0075Q-0.00098Q^2)\,x\\
\label{e7}
&&+(47.65-0.935Q+0.00743Q^2)\,x^2\:\mu\textrm{b/MeV}^2\:,
\end{eqnarray}
where $x=(m_{\pi\pi}-2m_{\pi})/m_{\pi}$.

Due to small recoil corrections, this parameterisation should be
used at an excess energy of $Q'$, where
\begin{eqnarray}\nonumber
Q'\approx xm_{\pi}+(Q-xm_{\pi})(1+2m_{\pi}/m_{\alpha})/(1+2m_{\pi}/m_p).\\
\end{eqnarray}

Since large Fermi momenta are not required in the two--step model,
the $dp\to\,^{3}\textrm{He}\,\pi^0$ and $\pi N\to \pi\pi N$
amplitudes can be taken outside of the integration in
eq.~(\ref{full}) with the values pertaining at zero Fermi momenta.
Considering only the positive energy pion pole, to first order in
$\bmath{k}$ and $\bmath{q}$ one is left with a difference between
the external and internal energies of
\begin{equation}
\Delta E=E_{\rm ext}-E_{\rm int}= \Delta
E_{0}+\bmath{k}\cdot\bmath{W}+\bmath{q}\cdot\bmath{V}\:,
\end{equation}
where
\begin{equation}
\Delta E_0 = E_{\pi}^0 -E_{\pi}\:,
\end{equation}
with
\begin{equation}
E_{\pi}^0=2E_d-E_t-E_n -E_{\pi}\:.
\end{equation}
Here ($E_{\pi},\,E_d,\,E_t,\,E_n$) are the pion, deuteron, triton,
and nucleon total energies, evaluated respectively at momenta
$-\fmn{3}{4} \bmath{p}_{\alpha}-\fmn{1}{2}\bmath{p}_d$,
$\bmath{p}_d$, $\fmn{1}{2}\bmath{p}_d$, and
$\fmn{3}{4}\bmath{p}_{\alpha}$. \vspace{1mm}

The relativistic relative velocity vectors $\bmath{V}$ and
$\bmath{W}$ depend only upon external kinematic variables:
\begin{eqnarray}
\nonumber \bmath{V} &=& \bmath{v}_{\pi}(-\fmn{3}{4}
\bmath{p}_{\alpha}-\fmn{1}{2}\bmath{p}_d)
- \bmath{v}_{n}(\fmn{1}{2}\bmath{p}_d)\\[1ex]
&=&-\frac{3}{4E_{\pi}}
\,\bmath{p}_{\alpha}-\frac{1}{2}\left[\frac{1}{E_{\pi}}
+\frac{1}{E_{n}}\right]\bmath{p}_d\:,
\nonumber \\[1ex] \nonumber
\bmath{W}&=&-\bmath{v}_{\pi}(-\fmn{3}{4} \bmath{p}_{\alpha}
-\fmn{1}{2}\bmath{p}_d)
+ \bmath{v}_{t}(\fmn{3}{4}\bmath{p}_{\alpha})\\[1ex]
&=&\frac{3}{4}\left[\frac{1}{E_{\pi}}+\frac{1}{E_t}\right]
\bmath{p}_{\alpha}+\frac{1}{2E_{\pi}} \,\bmath{p}_d\:.
\label{VandW}
\end{eqnarray}
The resulting form factor
\begin{eqnarray}
\nonumber%
\lefteqn{\mathcal{S}(\bmath{V},\bmath{W},\Delta
E_0)=}\\%
&&\hspace{-5mm}-i\int{\rd^3k} \,{\rd^3q}\,\frac{1}{\Delta
E_0+\bmath{k}\cdot\bmath{W} + \bmath{q}\cdot\bmath{V}
+i\epsilon}\,\tilde{\psi}^{*}(\bmath{k})\,
\tilde{\varphi}(\bmath{q})\nonumber\\
&=&(2\pi)^3\int_0^{\infty}\rd t \,e^{it\Delta E_0} \;
\psi^{*}(-t\bmath{W})\,\varphi(t\bmath{V})\:, \label{FF0}
\end{eqnarray}
then involves a one--dimensional integration over wave functions
in configuration space. In terms of this form factor the \ddpipi\
differential cross section becomes:
\begin{eqnarray}
\nonumber \rd\sigma&=&\frac{N_{\alpha}}{48\,(2\pi)^{10}}
\frac{p_{\alpha}p_d}{[m_pW(E_{\pi}+E_{\pi}^0)]^2}\,
|a(m_{\pi\pi},Q)|^2\,\times\\
&& \left|\mathcal{S}(\bmath{V},\bmath{W},\Delta E_0)+ (\bmath{p}_d
\Leftrightarrow -\bmath{p}_d)\right|^2\times \nonumber\\
&&\vphantom{\int}
\left\{\mid\mathcal{A}\mid^2+2\mid\mathcal{B}\mid^2\right\}\,
k_{\pi\pi}\,\rd{}m_{\pi\pi} \rd\Omega_{\alpha}\:,
  \label{simple}
\end{eqnarray}
where $N_{\alpha}$ is the normalisation of the $^4$He wave
function and the extra form--factor contribution coming from the
interchange of the two incident deuterons is indicated. All
isospin factors have been included, but it must be stressed that
in eq.~(\ref{simple}) $\mathcal{A}$ and $\mathcal{B}$ refer to the
$dp\to\,^{3}\textrm{He}\,\pi^0$ and $\pi^-p\to \pi^0\pi^0n$ charge
states respectively. Isospin invariance dictates that $\pi^+\pi^-$
production in \ddpipi\ should be a factor of two larger than
$\pi^0\pi^0$, but this simple rule is significantly modified near
threshold by phase--space factors arising from the pion mass
difference.

Two further refinements need to be implemented in
eq.~(\ref{simple}) before comparing its predictions with
experiment. Although the final $\alpha$-particle is slow in the cm
system, relativistic corrections cannot be neglected for the
incident deuterons. These can be included by boosting
$V_{\parallel}$, the longitudinal component of $\bmath{V}$,
\emph{i.e} by taking as argument of the deuteron wave
function~\cite{FW2}
\begin{equation}
\bmath{V}'=(\bmath{V}_{\!\perp},\,E_d V_{\parallel}/m_d)\:.
  \label{V-prime-def}
\end{equation}

Secondly, the effects of the deuteron $D$--state have to be
considered and this can be accomplished by introducing two form
factors:
\begin{eqnarray}
\nonumber%
\lefteqn{\mathcal{S}_{S,D}({V'},{W},\Delta E_0)}\\
&&=2\pi^2\!\!\int_0^{\infty}\rd t \,e^{it\Delta E_0} \,
\Psi^{*}(-t{W})\,\Phi_{S,D}(t{V'})\:,\label{FF1}
\end{eqnarray}
where $\Phi_{S,D}(r)$ are the deuteron $S$-- and $D$--state
configuration space wave functions normalised by
\begin{equation}
\int_{0}^{\infty}r^2\,\left\{\Phi_S(r)^2+\Phi_D(r)^2\right\}\,\rd
r =1\:.
\end{equation}
The $S$-- and $D$--state form factors enter in different
combinations for the $\mathcal{A}$ and $\mathcal{B}$ amplitudes
and, after making kinematic approximations in respect of the
$D$--state combined with the Lorentz boost, one finds
\begin{eqnarray}
\nonumber \lefteqn{\rd\sigma=\frac{N_{\alpha}}{48\,(2\pi)^{10}}
\frac{p_{\alpha}p_d}{[m_pW(E_{\pi}+E_{\pi}^0)]^2}\,
|a(m_{\pi\pi},Q)|^2\,k_{\pi\pi}}\\
&& \times
\left\{\mid\mathcal{A}\mid^2\left|\mathcal{S}_S({V'},{W},\Delta
E_0)-\sqrt{2}\,\mathcal{S}_D({V'},{W},\Delta E_0)
\right|^2\right. \nonumber\\
&& \nonumber\left.
+2\mid\mathcal{B}\mid^2\left|\mathcal{S}_S({V'},{W},\Delta
E_0)+\frac{1}{\sqrt{2}}\,\mathcal{S}_D({V'},{W},\Delta
E_0)\right|^2 \right\}\,\\
&&\hspace{5cm}\times\,\rd{}m_{\pi\pi}\, \rd\Omega_{\alpha}\:,
  \label{complex}
\end{eqnarray}
where, as in eq.~(\ref{simple}), it is assumed that contributions
from form factors resulting from the interchange $\bmath{p}_d
\Leftrightarrow -\bmath{p}_d$ have been included.

%
%
\newpage
\section{The $\mathbf{^4}$He wave function}
\label{sec3}%
\begin{figure}[htb]
\begin{center}
\centerline{\epsfxsize=8cm{\epsfbox{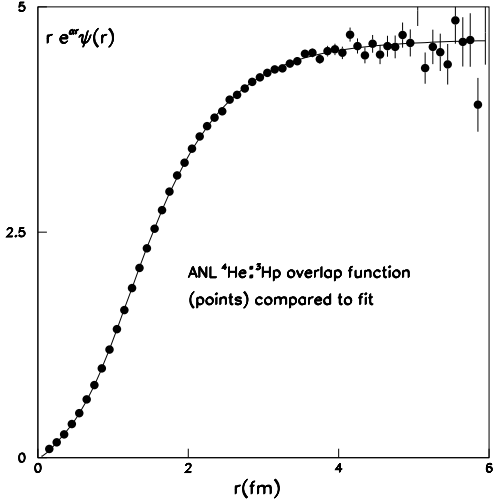}}}
\end{center}
\vspace{-1cm} \caption{\label{ANL} Unnormalised
$^4\textrm{He}\,$:$\,^3\textrm{H}\,p$ overlap function as a
function of the $^3$H--$p$ separation distance. For the purposes
of presentation, this has been multiplied by
$r\,\textrm{e}^{\alpha r}$, where the charge average
$\alpha=0.854$~fm$^{-1}$. The results of Ref.~\cite{VMC} have been
parameterised as in eq.~(\ref{param}).}
\end{figure}

Over the last few years there has been remarkable progress in
\emph{ab initio} calculations of the structure of light nuclei
using variational Monte Carlo techniques~\cite{VMC}. Starting from
realistic nucleon--nucleon potentials, it has been possible to
identify various cluster sub-structures in nuclei as heavy as
$^9$Be. The results for the unnormalised
$^4\textrm{He}\,$:$\,^3\textrm{H}\,p$ overlap function in
configuration space are shown in Fig.~\ref{ANL}, where the error
bars arise from the sampling procedure.\vspace{-10mm}

The overlap function has been parameterised by
\begin{equation}
\label{param} \psi(r)=\sqrt{N_{\!\alpha}}\,\frac{1}{r}
\sum_{n=1}^{6}a_n\,\textrm{e}^{-n\alpha r}\:,
\end{equation}
where $\alpha=0.854$~fm$^{-1}$ represents the average for the
$^3$H$\,p$ and $^3$He$\,n$ configurations. To ensure good
behaviour at the origin, the final parameter is fixed by
$a_6=-\sum_{n=0}^{5}a_n$, while the other values are sequentially
$5.1525$, $-2.8414$, $-45.1886$, $110.7401$, and $-100.3994$. The
normalisation has been chosen such that
\begin{equation}
\int_{0}^{\infty}r^2\,\left[\psi(r)\right]^2\,\rd r =
N_{\alpha}\:.
\end{equation}
In the spirit of our approach here to pion production, where only
these cluster contributions are considered, it is appropriate to
assume that the $p\,^3$H and $n\,^3$He components saturate the
wave function and take $N_{\,\alpha}=4$ rather than the reduced
spectroscopic factor obtained in ref.~\cite{VMC}.
%
%
\section{Results and Conclusions}
\label{Results}

\begin{figure}[htb]
\begin{center}
\centerline{\epsfxsize=8cm{\epsfbox{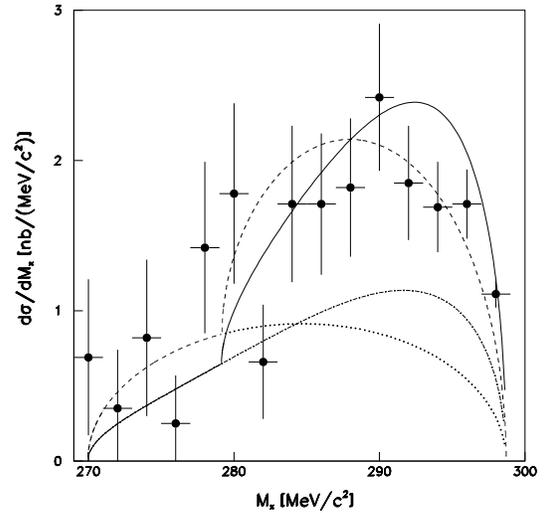}}}
\end{center}
\vspace{-1cm} \caption{\label{piafig} Missing--mass distribution
for the $dd\to\,^{4\,}\textrm{He}\,X$ reaction measured at
570$\:$MeV~\cite{Pia1}. The chain curve corresponds to
$\pi^0\pi^0$ production within the two--step model whereas the
solid one represents the sum of this and $\pi^+\pi^-$ production.
The predictions are normalised to the integrated cross section by
multiplying by a factor of 17.6. The dotted and broken curves are
the similar predictions from phase space, again normalised to the
total rate.}
\end{figure}

In Fig.~\ref{piafig} we show the prediction of the shape of the
missing--mass distribution for inclusive two--pion production at
an excess energy of $Q=29\:$MeV with respect to the $2\pi^0$
threshold. Though the general form is in good agreement with the
experimental data~\cite{Pia1,Pia2}, the results are too low by
almost a factor of twenty! The peak of the distribution is
predicted to be a little to the right of that corresponding to
pure phase space, which is also shown. Such a feature was clearly
observed for the \pdpipi\ reaction at low energies~\cite{FGW}, but
the limited statistics in the $dd$ case prevents us from drawing
firm conclusions here. Estimates in the double--$\Delta$
model~\cite{Anders}, which agreed convincingly with the data in
the resonance region, were even poorer compared to the
near--threshold data. Apart from being a similar factor of twenty
too low, this model also predicted significant structure in the
mass distribution which is absent from the experimental
data.\vspace{-5mm}

\begin{figure}[htb]
\begin{center}
\centerline{\epsfxsize=8cm{\epsfbox{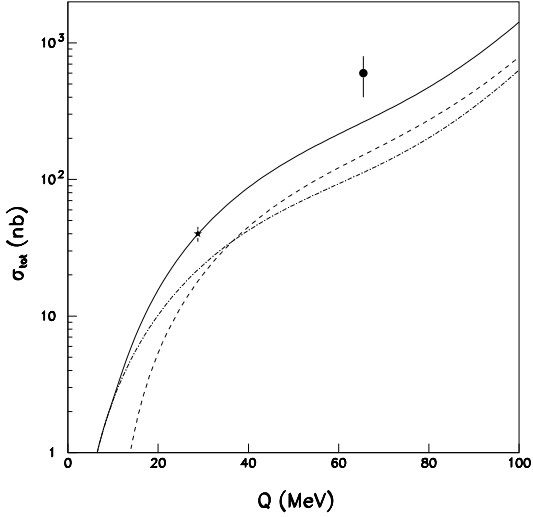}}}
\end{center}
\vspace{-1cm} \caption{\label{sigtotpia} Total cross section for
the \ddpipi\ reaction. The experimental data from Ref.~\cite{Pia1}
(star) and Ref.~\cite{Chapman} (circle) are compared to the
predictions of the two--step model scaled by a factor of $17.6$.
The chain curve corresponds to $\pi^0\pi^0$ production, the broken
to $\pi^+\pi^-$, and the solid to their sum. }
\end{figure}

The discrepancy is similar for the other low energy
data~\cite{Chapman}, though here the acceptance was small and
assumptions had to be made in order to extract a total cross
section. In Fig.~\ref{sigtotpia} we show the estimates of the
total cross sections for the production of charged and neutral
pions within the two--step model.

The central problem for any model that attempts to describe the
\ddpipi\ cross section at low energies is that the production of
isoscalar pion pairs is very similar in deuteron--deuteron and
proton--deuteron collisions. Thus at
$Q=29\:$MeV~\cite{Andersson,Pia1},
\begin{equation}
\frac{\sigma_{\rm tot}(\ddpipi)}{\sigma_{\rm tot}(\pdpipi)}\approx
\frac{40\:\textrm{nb}}{60\:\textrm{nb}}=\frac{2}{3}\:\cdot
\end{equation}

On the other hand, the production of the $\eta$ meson is much
weaker in the $dd$ case, with the ratio of the squares of the
amplitudes being~\cite{Frascaria,Willis,Berger2,Mayer}
\begin{equation}
\frac{\left|f(\ddeta)\right|^2}{\left|f(\pdeta)\right|^2} \approx
\frac{1}{50}\,,
\end{equation}
though perhaps this would be increased by a factor of two if
corrections were made for the effects of the $\eta$--nucleus
final--state interaction. Since the low energy \pdpipi, \pdeta,
and \ddeta\ cross sections are all successfully described by the
two--step model, a factor of ten undershoot in the \ddpipi\ case
is not too surprising. The crude comparison made here does not
take into account fully the spin--parity considerations and the
prediction would have been increased by more than a factor of two
if the sign of the deuteron $D$--wave were reversed in
eq.~(\ref{complex}).

Given that neither the two--step nor the double--$\Delta$ model
seems capable of describing the magnitude of the \ddpipi\ cross
section near threshold, one must seek alternative explanations or
modifications to the existing mechanisms. Other diagrams, such as
that of the impulse approximation where the process is driven by
\pdpipi\ with a spectator nucleon, give very small cross sections
due to the large momentum transfer. We have not included any
specific $\pi\pi$ final--state interaction, but the $s$--wave
scattering lengths are relatively small~\cite{Batley} and, in any
case, the effects are implicitly included through the use of
empirical $\pi N\to \pi\pi N$ amplitudes~\cite{Oset}.

The interaction of the low energy pions with the final $^4$He
nucleus might enhance the cross section since it is known that the
$p$--wave pion--nucleus interaction is attractive near
threshold~\cite{Ericson}. However, the effect will steadily
diminish with energy and eventually change sign at the resonance.
Crude estimates indicate that any effects due to such final--state
interactions are likely to be less than 50\%, even very close to
threshold, and so they are very unlikely to provide the
explanation of the defect.

Now, although we have normalised the $^4$He wave function as if it
consisted purely of $p\,^3$H/$n\,^3$He pairs, in reality the $^3$H
in such a nucleus is on average smaller than the physical triton.
Nevertheless we have taken the amplitudes for
$pd\to\,^{3}\textrm{He}\,\pi^0$ from the measured data. The same
criticism can be levelled at the double--$\Delta$ model, where the
final deuteron in the $pp\to d\pi^+$ input would really be
required for a \emph{small} deuteron. If there were major
corrections due to such effects they would be likely to be present
at all energies and hence destroy the excellent agreement with
data achieved at higher energies~\cite{Anders}. Further
inspiration is therefore clearly needed to resolve this dilemma.

%
%
\begin{acknowledgement}
This work has been much influenced by long--standing discussions
with Pia Th\"orngren, which have been beneficial to both sides.
One of the authors (CW) is appreciative of the hospitality shown
to him by Uppsala University. Support from the EtaNet programme of
the EU is gratefully acknowledged.
\end{acknowledgement}

%
%

\end{document}